# Interplay of waves and eddies in rotating stratified turbulence and the link with kinetic-potential energy partition

Raffaele Marino[1,2,3], Duane Rosenberg[4,5], Corentin Herbert[6] and Annick Pouquet[7,8]
[1]*École Normale Supérieure de Lyon,*
*Université de Lyon, F-69007 Lyon, France.*
[2]*Dipartimento di Fisica,*
*Università della Calabria, 87036 Rende, Italy.*
[3]*Space Sciences Laboratory,*
*University of California Berkeley,*
*Berkeley, CA 94720, USA.*
[4]*National Center for Computational Sciences,*
*Oak Ridge National Laboratory,*
*Oak Ridge, TN 37831, USA.*
[5]*SciTec, Inc., Princeton, NJ 08540, USA.*
[6]*Department of Physics of Complex Systems,*
*Weizmann Institute of Science, Rehovot 76100, Israel.*
[7]*Laboratory for Atmospheric and Space Physics,*
*University of Colorado, Boulder, CO 80309, USA.*
[8]*National Center for Atmospheric Research,*
*Boulder, CO 80307, USA.*

The interplay between waves and eddies in stably stratified rotating flows is investigated by means of world-class direct numerical simulations using up to $3072^3$ grid points. Strikingly, we find that the shift from vortex to wave dominated dynamics occurs at a wavenumber $k_R$ which does not depend on Reynolds number, suggesting that partition of energy between wave and vortical modes is not sensitive to the development of turbulence at the smaller scales. We also show that $k_R$ is comparable to the wavenumber at which exchanges between kinetic and potential modes stabilize at close to equipartition, emphasizing the role of potential energy, as conjectured in the atmosphere and the oceans. Moreover, $k_R$ varies as the inverse of the Froude number as explained by the scaling prediction proposed, consistent with recent observations and modeling of the Mesosphere–Lower Thermosphere and of the ocean.

## I. INTRODUCTION

In the Planetary Boundary Layer (PBL), waves together with the occurrence of strong bursts of activity are often observed at night, when the PBL is more stably stratified. The origin of this feature is not well understood [1], but it has been shown recently that, for strong enough stratification, the waves can increase the development of negative vertical velocity gradients in an interval of Froude numbers [2], a feature which can be associated with critical layers [1, 3]. Waves also play an essential role in the exchanges between the ocean and the atmosphere, and thus in climate dynamics [4] since they enhance mixing and modify heat and mass fluxes.

Thus, it is of major importance to study how the partition of energy between waves and slow (vortical, quasi-geostrophic or QG) modes vary with relevant parameters of rotating stratified turbulence (RST) and what are the underlying physical mechanisms. Different regimes occur which can be described by reduced equations obtained utilizing a small parameter, the ratio of the wave period to the nonlinear turn-over time $\tau_{NL}$ [5, 6]. Atmospheric observations suggest that the slow and fast mode energies are comparable at mesoscales [7, 8], with an important role played by potential energy [9, 10], and one finds that the rotational component (based on vertical vorticity) dominates over the divergent one due to gravity waves at large scale, up to roughly 400 $km$, corresponding to the synoptic to meso-scale transition, the latter being also observed in the ocean [11]. In fact, it is known that the direct cascade of energy to small scales is due in part to nonlinear triadic interactions between eddies and waves [12–15], but that waves and vortices exchange less energy in the presence of rotation than for purely stratified flows [12]. Moreover, the total energy is now being transferred both to the large and to the small scales with a dual constant-flux energy cascade [16, 17], as also recently observed for the ocean using sea-surface heights measurements [18, 19]. Thus, this wave-vortex transition is quite a general feature of atmospheric and oceanic flows although there is no clear consensus as to what governs the scale at which it occurs.

There is a number of laboratory experiments examining the decay of energy in RST. For example, the study in [20], which spans a large range in Reynolds numbers, deals with turbulence generated by a set of vertical plates, corresponding to a quasi-two-dimensional (2D) field, and the resulting flows seem to be well described by the QG approximation. A strong slowing-down of the energy decay in the presence of rotation is observed, as expected for wave turbulence; it is mainly attributed to the enstrophy cascade of the QG regime, with a horizontal energy spectrum $\sim k_\perp^{-3}$ [20].

In this paper we show for the first time, in the Boussinesq framework, that a simple scaling emerges for the wavenumber $k_R$ at which the shift occurs from a vortex-dominated to a wave-dominated dynamics in rotating stratified turbulence, and that such transitional wavenumber does not depend on the Reynolds number. In fact, here the transition between these regimes proves to be mainly controlled by stratification while it is only weakly or not affected by rotation and the development of turbulence at the smaller scales.

## II. FRAMEWORK FOR THE PARAMETRIC STUDY

Four dimensionless parameters govern the behavior of RST, namely the Reynolds, Froude, Rossby and Prandtl numbers: $Re = U_0 L_0/\nu$, $Fr = U_0/[L_0 N]$, $Ro = U_0/[L_0 f]$, $Pr = \nu/\kappa$, where $U_0$, $L_0$ are characteristic velocity and length and $\nu$, $\kappa$ are the viscosity and diffusivity taken equal ($Pr = 1$), with $\mathcal{R}_B = ReFr^2$ the buoyancy Reynolds number.

Energy partition between the wave and slow modes has been studied numerically in [14] in the context of RST in a cubic box, with large-scale white noise in time forcing. Defining $\epsilon_{SM} = f/N$, a change is found for $\epsilon_{SM} \approx 1$, from a wave-dominated to a slow-mode (QG) regime as $\epsilon_{SM}$ decreases. In [21, 22] similar computations are performed at high resolution for unit Burger number, $Bu = NH/[fL]$ where H and L are vertical and horizontal length scales. Runs therein were restricted to cases when the potential vorticity (PV) remains linear and for equal strength of rotation and stratification, with $Fr \approx 0.002$; thus, the buoyancy scale $L_B = U_0/N$ is not resolved and $\mathcal{R}_B < 1$; indeed, even if based on a (high) equivalent Reynolds number using hyper-diffusivities, $\mathcal{R}_B > 1$ requires the Ozmidov scale $\ell_{Oz} \sim Fr^{1/2} L_B$ to be resolved, $\ell_{Oz}$ being the scale at which isotropy and Kolmogorov scaling recover for stratified flows. This choice of parameters corresponds to the regimes attainable by asymptotic analyses of RST (see e.g. [5, 6]). Under these conditions, the vortical energy becomes dominant as $N/f$ increases [22].

In this paper, in contrast, we examine flows in the absence of forcing and hyper-diffusion, without resorting to the use of models for the small scales, in regimes with specifically $\mathcal{R}_B$ up to $1.28 \times 10^5$ and $Re$ up to $5.4 \times 10^4$. If these parameters remain small compared to the troposphere or the ocean, the Mesosphere–Lower Thermosphere (MLT) provides an exception [23], and runs in a parameter space realistically compatible with this region are considered as well in this numerical exploration.

Denoting **u** the incompressible velocity field, $\nabla \cdot \mathbf{u} = 0$, and $\theta$ the temperature fluctuations, the Boussinesq equations are:

$$\frac{\partial \mathbf{u}}{\partial t} + N\theta\hat{z} + f\hat{z} \times \mathbf{u} - \nu\nabla^2 \mathbf{u} = -\nabla p - \mathbf{u} \cdot \nabla \mathbf{u} \quad (1)$$

$$\frac{\partial \theta}{\partial t} - N\mathbf{u} \cdot \hat{z} - \kappa\nabla^2 \theta = -\mathbf{u} \cdot \nabla \theta \; ; \quad (2)$$

$N = \sqrt{-g\partial_z\bar{\theta}/\theta_0}$ with $\partial_z\bar{\theta}$ the background imposed stratification, and $p$ is the pressure normalized to a unit mass density. We use the pseudo-spectral code GHOST (Geophysical High-Order Suite for Turbulence) which implements triply periodic boundary conditions in an adimensionalized cube of size $2\pi$, on a grid of $n_p^3$ points; here, $n_p = 1024$ for 48 runs and three more runs were performed with $n_p = 512, 2048$ and $3072$ respectively. The code is parallelized with a hybrid method [24], and a second-order Runge-Kutta scheme is employed. Dimensionless parameters and other quantities introduced hereafter are measured at peak of dissipation, with specifically $L_0$ the integral scale and $U_0$ the $r.m.s.$ velocity. We tested that all the runs were sufficiently resolved insofar as $R_{res} = k_{max}/k_\eta \geq 1$, where $k_{max} = n_p/3$, $k_\eta = [\epsilon_T/\nu^3]^{1/4}$ being the dissipation wavenumber for a Kolmogorov spectrum with $\epsilon_T = |DE_T/Dt|$ the total energy dissipation rate. Initial conditions for the velocity are random, isotropic in Fourier space and centered on $2\pi/k_0$ in the large scales ($2 \leq k_0 \leq 3$), while the initial potential energy is zero.

Visualization of the total vorticity magnitude $|\omega|$ at the peak of dissipation for a MLT run with $N/f = 137, Re = 11728$ and $f = 0.04$ is given in Fig. 1, the vorticity being defined as usual as $\omega = \nabla \times \mathbf{u}$ with **u** the velocity field. One can see that the local micro-Rossby number $\omega(\mathbf{x})/f$ is quite high everywhere in the flow and so is the Rossby number for that flow. Traces of the stratified layers are noticeable as well, together with intense activity which is linked with local instabilities.

In absence of nonlinear terms, the Boussinesq equations can support waves, and the eigenmodes are determined in a straightforward manner (see e.g. [12, 25–27]). This decomposition is used to build the phase space of the underlying dynamical system. It is shown in [28] to be instrumental in unraveling the role of the slow modes in preventing the inverse cascade of energy in stratified flows [29, 30] when using statistical ensembles restricted to the linear formulation of PV.

Summarizing the expression of spectra in terms of linear modes, the basic physical fields, transformed into Fourier space for each wave vector **k** (of modulus $k$), can be regrouped as $\mathbf{X}(\mathbf{k}) = [u_i(\mathbf{k}), \theta(\mathbf{k})], i = 1,3$. Using incompressibility, they are decomposed onto the eigenmodes of the linearized Boussinesq equations in terms of one slow ($\mathbf{X}_0(\mathbf{k})$) and two fast wave modes ($\mathbf{X}_\pm(\mathbf{k})$), with amplitudes $A_{0,\pm}(\mathbf{k})$ as:

$$\mathbf{X}(\mathbf{k}) = A_0(\mathbf{k})\mathbf{X}_0(\mathbf{k}) + A_+(\mathbf{k})\mathbf{X}_+(\mathbf{k}) + A_-(\mathbf{k})\mathbf{X}_-(\mathbf{k}) \;.$$

The energy Fourier spectra and their sums are written as usual as $E_0 = \langle u_0^2 + \theta_0^2 \rangle/2 = \Sigma_\mathbf{k}|A_0(\mathbf{k})|^2 = \Sigma_\mathbf{k}E_0(\mathbf{k})$

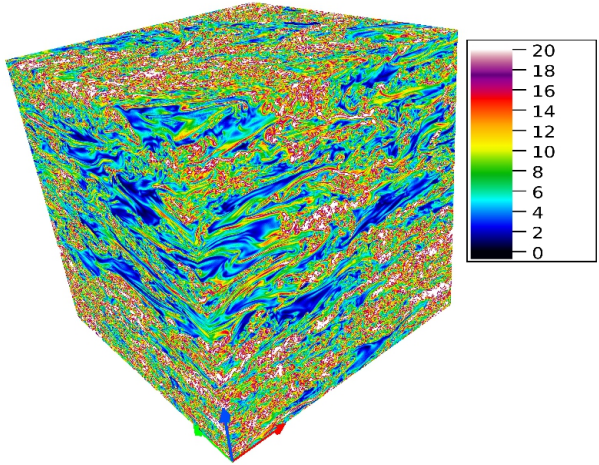

FIG. 1. Visualization of the magnitude of total vorticity, using a linear color bar, for a DNS flow with parameters realistically compatible to those of the MLT region in the atmosphere: $N/f = 137$, $Fr = 0.067$, $Ro = 9.2$, $Re \simeq 12000$ and $\mathcal{R}_B = 53$. The common direction of gravity and rotation is indicated by the blue arrow.

and $E_W = \langle u_w^2 + \theta_w^2 \rangle /2 = \Sigma_{\mathbf{k}} E_W(\mathbf{k})$, with (see [31]):

$$E_W = \Sigma_{\mathbf{k}} |A_+(\mathbf{k})|^2 + |A_+(\mathbf{k})|^2 = \Sigma_{\mathbf{k}} \left[ E_+(\mathbf{k}) + E_+(\mathbf{k}) \right] .$$

## III. PARTITION BETWEEN WAVE AND SLOW MODES

Fig. 2 (top) gives the wave and slow-mode isotropic spectra of the total (kinetic plus potential) energy $E_T = E_V + E_P$ for runs with $N/f = 4.96, Fr = 0.025$, and with Reynolds numbers spanning the range $2600 \leq Re \leq 54000$ using different grids from $512^3$ to $3072^3$ points. For all, they cross at $k_R \approx 7$ (note that wavenumbers being integers, $k_R$ is determined as the closest wavenumber after crossing, or after the last crossover, highest in wavenumber, when spectra cross at more than one point.) A striking result is that $k_R$ does not show dependence on Reynolds number: it is not determined by how vigorous (and turbulent) the small-scale dynamics is, as measured by $Re$ (or $\mathcal{R}_B$), but rather represents multi-scale interactions between slow and fast modes. Note that the ratios $R_{res}$ are also indicated in the figure and they fall in the range $1.04 \leq R_{res} < 2.33$, thus showing that numerical resolution as measured by $R_{res}$ does not modify the evaluation of $k_R$.

The large-scale dominance of slow modes is associated with the role of rotation leading, in the presence of forcing, to an inverse vortical energy cascade, whereas gravity modes take over at small scales through nonlinear coupling, in the weak and strong turbulence regimes [12]. This is compatible with a large-scale QG regime and with the fact that the inertia-gravity wave frequencies are limited to the range $f \leq \sigma_k \leq N$ for $N/f > 1$ as in this study. In other words, the fastest waves are associated with stratification and they are the ones which can compete with nonlinear eddies in a wider range of (small-scale) wave numbers.

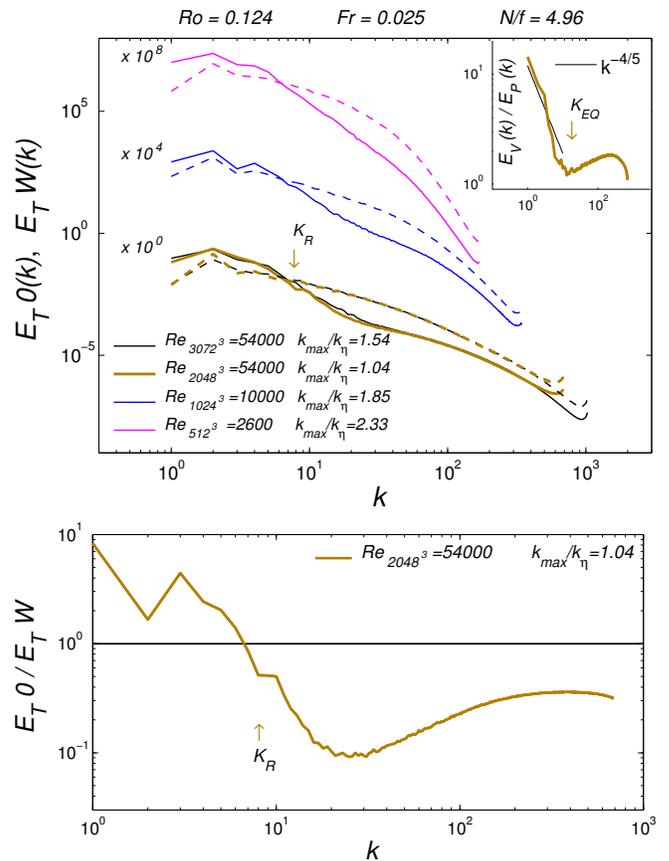

FIG. 2. (Top): Slow and wave-mode (solid and dashed lines) total isotropic energy spectra at peak of dissipation, respectively $E_T 0(k)$ and $E_T W(k)$, for runs with identical $Ro$ and $Fr$ but different Reynolds number and resolution; all cross at $k_R \approx 7$. *Insert*: ratio of kinetic to potential energy spectra, minimum at $k = k_{EQ}$, for the run with $Re = 54000$, $n_p = 2048$. The wavenumber $k_{EQ}$ is comparable to $k_R$ and the slope of the ratio $E_V(k)/E_P(k)$ is compatible with that predicted in the Bolgiano-Obukhov phenomenology. (Bottom:) Ratio of slow to wave mode total isotropic energy spectra for the same run as in the insert, on a grid of $2048^3$ points.

The upper insert in Fig. 2 (top) shows the ratio of kinetic to potential energy spectra, $\rho_E(k) = E_V(k)/E_P(k)$, for the run in the main plot with $n_P = 2048^3$: it has a sharp decrease at large scales $\approx k^{-4/5}$, a scaling compatible with a Bolgiano-Obukhov (BO) law [32, 33] due to the determining role of potential energy as the source in a nonlinear turbulent cascade (see also [34] where a BO range is clearly identified).

The decrease of kinetic energy with scales leads to a shortening of the eddy turn-over time $\tau_{NL} = [k^3 E_V(k)]^{-1/2}$ (provided the energy spectrum is not steeper than $k^{-3}$); eventually, the nonlinear eddies become faster than the waves and can compete efficiently with them. Thus, at the end of this range, another regime arises which corresponds to strong nonlinear interactions.

In the insert, $k_{EQ}$ denotes the wavenumber at which the lowest value of $\rho_E$ is reached: $\chi_{EQ} = \rho_E(k_{EQ}) = min[\rho_E]$. The scale $L_R = 2\pi/k_R$ is comparable to the characteristic Bolgiano-Obukhov scale in the cases where the BO scaling prediction is fulfilled for kinetic and potential energy spectra, as for instance in the high resolution runs of Fig.2. Once the large-scale dynamics has settled, a scale-by-scale quasi-equipartition is obtained, with $\rho_E(k) \approx 2$, in agreement with atmospheric data in the mesoscales [35]. We also find a rather slow growth of $\rho_E(k)$ as $k$ increases. Note that equipartition is expected in the regime in which waves prevail, but that it is also a common feature of turbulent flows at small scale. It is linked to the statistical properties of the flows that depend on the ideal invariants, as in MHD with equipartition between kinetic and magnetic energy (see e.g. [36]). Another feature of these spectra is that the kinetic (not shown) and total energies behave in similar ways, with a steady decrease of the wave-to-slow mode ratio as one moves to smaller scales. Finally, we show in Fig. 2 (bottom) the ratio of the slow-mode to wave-mode total energy spectra for the run on a grid of $2048^3$ points with $\mathcal{R}_B \approx 32$. The change of behavior for the wavenumber $k_r \approx 20$ may be related to the role of rotation relative to that of stratification, with $k_r$ close to the wavenumber corresponding to the radius of deformation when computed using $k_0$, namely $k_D = 2\pi/L_D = [N/f]k_0$, with here $k_0 = 2.5$ and $N/f \approx 5$.

Using the fact that $k_R$ is independent of $Re$, a parametric study is now analyzed with $n_p = 1024$ and $0.002 < Fr < 5.5$, $0.1 < Ro < 42$, $2.5 < N/f < 312$, $2520 < Re < 5.4 \times 10^4$, $0.053 < \mathcal{R}_B < 1.28 \times 10^5$, and $1.02 < k_{max}/k_\eta < 2.58$. A few more runs are done at very large or infinite $Ro$ (indicated in Fig. 3, bottom).

We plot in Fig. 3 (top) $k_R$ as a function of $Fr$, finding a scaling compatible with $k_R \sim 1/Fr$, very clear for runs with $Ro < 0.5$ (filled symbols), and saturation at $k_R/k_0 \approx 1$ for high $Fr$. Estimated values of $k_R$ are binned here in Reynolds number; this further confirms the lack of dependency of the crossing wavenumber on small-scale dynamics. In fact, symbols corresponding to a same bin in $Re$ exhibit variations of one order of magnitude in $k_R/k_0$, thus corroborating (and complementing) the analysis reported in Fig. 2.

One can argue that the shift from QG in the large scales to weak turbulence (WT) at smaller scales depends on the small WT parameter, namely $Fr$ in a simple (linear) fashion. A $k_R \sim Fr^{-1}$ scaling is also compatible with the invariance of the Boussinesq equations as stated in [37], ensuring that the vertical Froude number be of order unity. As in Fig. 2, $k_R \sim k_{EQ}$ for these runs as shown in Fig. 3 (middle), where $k_{EQ}$ is as before the wavenumber at which the ratio $\rho_E(k) = E_V(k)/E_P(k)$ reaches its minimum $\chi_{EQ}$ (or second minimum when there are two). The insert in Fig. 3 (middle) gives the variation of $k_R/k_0$ as a function of $\chi_{EQ}$, showing a strong correlation between the two. In relative terms, as the Froude number decreases, $k_R/k_0$ increases since the influence of stratification becomes stronger through a larger range of scales; this corresponds, for given $U_0, L_0$, to a larger Brunt-Väisälä frequency, i.e. to faster waves relative to eddies. A change of regime will occur when a balance is achieved between linear waves and nonlinear advection. For flows with increasing stratification, the nonlinear eddies thus have to become faster as well for balance to be reached. Hence, kinetic energy has to increase further relative to potential energy which is the source of the wave dynamics in this regime. The scaling found in the insert of Fig. 3 (middle) is consistent with such a physical interpretation since $\tau_{NL}^{-1}$ is linearly dependent on velocity when evaluated locally at a given scale.

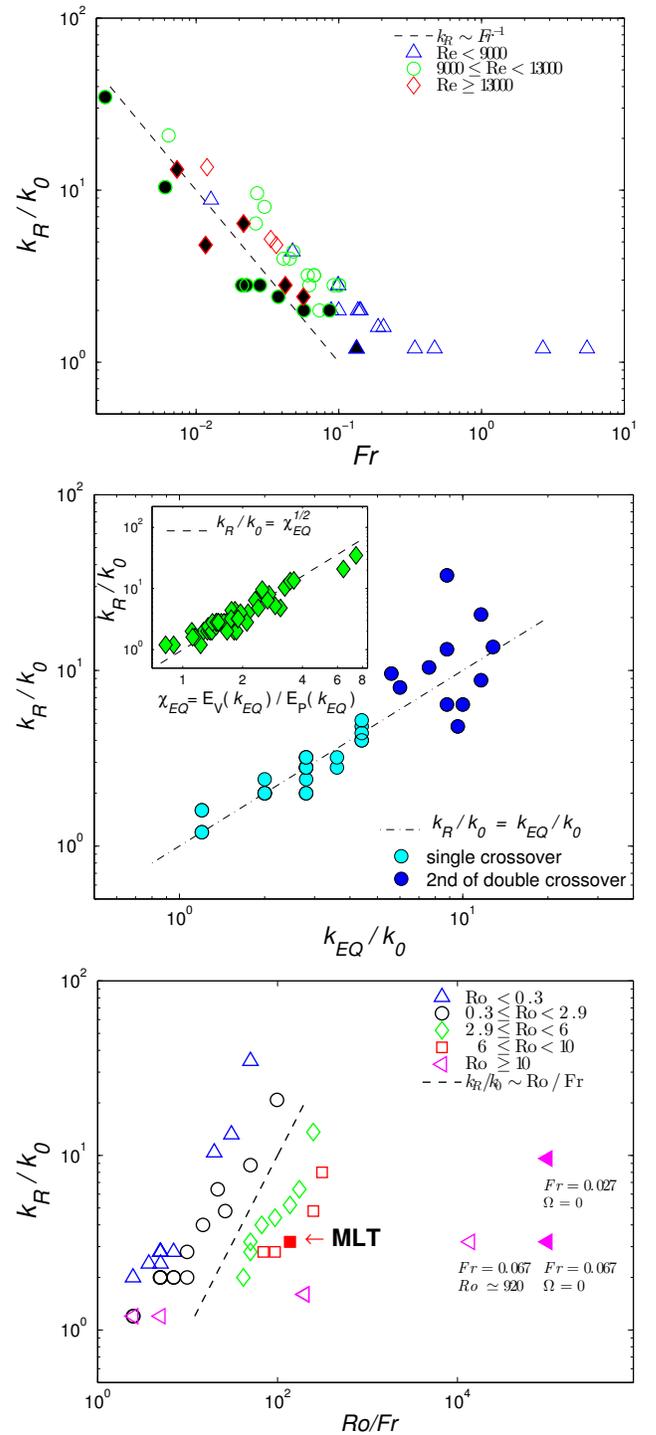

FIG. 3. (Top:) Normalized crossing wavenumber $k_R/k_0$ versus $Fr$ in log-log coordinates. Symbols correspond to different bins in Reynolds number with the filled points identifying runs characterized by $Ro < 0.5$. (Middle:) $k_R/k_0$ vs. $k_{EQ}/k_0$, with $k_{EQ}$ the wavenumber at which $E_V(k)/E_P(k)$ reaches its minimum value, $\chi_{EQ}$. In the insert is given $k_R/k_0$ as a function of $\chi_{EQ}$. (Bottom:) $k_R/k_0$ vs. $N/f$ binned in Rossby number. The line indicates approximate scaling. Filled (magenta) triangles are purely stratified runs whereas the filled red square refers to a run whose parameters are realistically compatible with that of the atmospheric MLT region. Dashed lines give scaling as indicated.

When examining now the effect of rotation in Fig. 3 (bottom) by plotting $k_R/k_0$ as a function of $N/f$, binned this time in Rossby number, the scaling in $1/Fr$ persists and changing $Ro$ simply shifts the data points with a saturation at high $Ro$. Indeed, it is known that there is a critical Rossby number $\approx 0.1 - 0.2$ above which the influence of rotation is greatly diminished, and in the forced case the inverse cascade of energy disappears. This point was also observed in [17] when expressed in terms of a change of behavior of the relative strength of the inverse and direct energy fluxes for $N/f \approx 7$, corresponding to a transition for $Ro \simeq 0.45$.

These results can be recast in the following way. At wavenumber $k_R$, the dominance of vortex modes over wave modes is arrested, a fact that can be attributed to numerous weakly nonlinear wave interactions. As for purely stratified flows, the thickness of layers that form in rotating stratified turbulence with large $N/f$ is determined by the generalization to RST of the buoyancy scale $L_B = FrL_0 \sim 1/N$ (a scale associated with a vertical Froude number of order unity [37]). As rotation increases, the stratified layers become slanted (see Fig. 1, and see renders in [38]) in a way that is compatible with an anisotropic equilibration between vertical and horizontal dynamics, namely $N/L_\perp \sim f/L_\parallel$, making in fact the two components of the dispersion relation of comparable magnitude and indicating a balance as occurs in the Rossby deformation radius, $R_D = [N/f]L_\parallel$ with however the vertical length scale not determined a priori but through the intrinsic dynamics of the flow [37]. These layers make an angle $\delta$ with respect to the vertical; one can assume that the vertical length scale now varies as $L_B/\sin\delta$ which, to lowest order, will be proportional to Ro, namely $L_\parallel \sim L_0[Fr/Ro]$, a generalization from the purely stratified case already hypothesized in [37] and which is confirmed by the present study. At the scale at which one switches from rotation-dominated (with strong kinetic energy and an inverse cascade in the forced case) to wave-wave interactions with quasi equipartition between $E_V$ and $E_P$ throughout a smaller-scale inertial range, the respective roles of eddy interactions and wave interactions shift as well.

Quasi-equipartition between $E_V$ and $E_P$ at small scales has also been observed in atmospheric data of the mesoscales [35] and in numerical simulations [10, 39–41]. Energetic exchanges between waves and vortices and between potential and kinetic modes occur at all scales, and among scales in rotating stratified flows, implying the necessity to resolve all characteristic scales. The waves keep a strong influence in the small scales due in part to the fact that vortical modes have no vertical velocity and yet one eventually expects isotropy and equipartition on average to recover, implying $E_W \approx E_0/2$ at these scales. The small-scale dominance of waves for large $Re$ may seem paradoxical but it is likely governed by local instabilities with strong emission of gravity waves, with a small local Richardson number in an increasing proportion of the overall flow as $Re$ grows (see [42]). Also, we note that already at a Froude number of 0.04, the Ozmidov scale $\ell_{Oz} \sim Fr^{3/2}L_0$ at which isotropic Kolmogorov scaling is supposed to be recovered is barely resolved in runs with $n_p = 1024$ and $k_0 = 2.5$. It is only for the runs with higher $Fr$ (and thus higher $\mathcal{R}_B$) that the waves begin to subside and the ratio $E_0(k)/E_W(k)$ gets closer to (but still below) unity in the small scales (not shown).

## IV. DISCUSSION

The role of potential energy in the dynamics of rotating stratified turbulence has been emphasized over the years, in particular in the context of more accurate sub-grid scale parametrizations of the nocturnal planetary boundary layer (see e.g. [43, 44]). The initial conditions of the runs analyzed in this paper all have velocity modes only, and they are centered in the large scales. We know that balanced initial conditions also give a similar scaling with an identical value of $k_{EQ}$ [34]. In the future, we plan to analyze the stationary case, using both isotropic and balanced forcing; this will also allow for longer time statistics.

It is a combination of observations and experiments, numerical simulations and modeling that will lead to progress, since exploration at higher buoyancy Reynolds numbers to enter regimes of closer interest to the great majority of oceanic and atmospheric systems, with the need to keep both the Froude and Rossby numbers small and yet $ReFr^2$ high, will remain a challenge for some time, although a significant step has been taken in this direction within the present paper. In this context, we note that recent atmospheric simulations show that the crossing between the wave and vortex modes occur at a smaller scale for the tropospheric (less stable) flow than for the stratospheric (more stable) flow [41], in agreement with our $k_R \sim Fr^{-1}$ result.